\documentclass[hyper]{JHEP} 

\usepackage{epsfig}

\newcommand\fverb{\setbox\pippobox=\hbox\bgroup\verb}

\newcommand\fverbdo{\egroup\medskip\noindent%

            \fbox{\unhbox\pippobox}\ }

\newcommand\fverbit{\egroup\item[\fbox{\unhbox\pippobox}]}

\newbox\pippobox

\title{Nonrelativistic String Theory Sigma Model and Its Canonical Formulation}
\author{J. Kluso\v{n}\\
Department of
Theoretical Physics and Astrophysics\\
Faculty of Science, Masaryk University\\
Kotl\'{a}\v{r}sk\'{a} 2, 611 37, Brno\\
Czech Republic\\
E-mail: \email{klu@physics.muni.cz}} \preprint{}

 \abstract{This paper is devoted to the canonical analysis
 of  non-linear sigma model that describes motion of non-relativistic
string on stringy Newton-Cartan background. We determine structure of constraints
of this string and compare resulting Hamiltonian  with previous proposal of non-relativistic string on
Newton-Cartan  background.}

\def\mM{\mathcal{M}}

\def\hE{\hat{E}}

\def\hE{\hat{E}}

\def\blambda{\bar{\lambda}}

\def\be{\begin{equation}}

\def\ee{\end{equation}}

\def\bea{\begin{eqnarray}}

\def\htau{\hat{\tau}}
\def\eea{\end{eqnarray}}

\def\mH{\mathcal{H}}

\newcommand{\mG}{\mathcal{G}}

\newcommand{\bT}{\mathbf{T}}

\newcommand{\mL}{\mathcal{L}}

\def\pb #1{\left\{#1\right\}}

\begin{document}
\section{Introduction and Summary}
In \cite{Gomis:2000bd,Danielsson:2000gi} non-relativistic string theory with Galilean invariant global symmetry
was proposed
\footnote{For related works, see for example
\cite{Gomis:2016zur,Batlle:2016iel,Sakaguchi:2007ba,Casalbuoni:2007rs,Kim:2007hb,Sakaguchi:2007zsa,Kluson:2006xi,Sakaguchi:2006pg,Brugues:2006yd,Gomis:2005bj,Gomis:2005pg,Gomis:2004ht,Gomis:2004pw,Brugues:2004an}.}. This theory is described by two-dimensional quantum field theory which is well defined and which contain fields that describe dynamics of string in target space-time together with additional fields which are crucial for consistency of string theory. It is important to stress that the target space-time,  where the string propagates corresponds to flat space-time invariant under Galilean symmetry. The characteristic property of non-relativistic string theory is that there is no Riemannian
 metric in the target space. In fact, non-relativistic string theory provides a quantization of non-relativistic space-time geometry in the same way as relativistic string theory provides quantization of Riemannian geometry. Natural question is whether we can formulate non-linear sigma model that describes string propagation on a non-relativistic target space-time structure. As was shown recently in \cite{Bergshoeff:2018yvt} the appropriate geometry corresponds to so-called stringy Newton-Cartan geometry
\cite{Andringa:2012uz}
\footnote{For related works, that analyze point particles or extended objects in Newton-Cartan geometry or its stringy generalizations, see for example
\cite{Kluson:2018egd,Kluson:2017abm,Barducci:2017mse,Kluson:2017pzr,Kluson:2017djw,Harmark:2017rpg,Kluson:2017vwp}.}.

The action proposed in \cite{Bergshoeff:2018yvt} is very interesting and certainly deserves further study. In particular, it would be very nice to find Hamiltonian form of this action and analyze its relation to Hamiltonian that was proposed recently in \cite{Kluson:2018uss}. The goal of this paper is to perform such an analysis. It turns out that the canonical analysis of the action proposed in \cite{Bergshoeff:2018yvt} is rather non-trivial due to the complicated structure of the target space-time and also thanks to the presence of additional world-sheet fields that are needed for the consistency of  theory. Since these fields are non-dynamical we find that  their conjugate momenta are the primary constraints of the theory. Then requirement of the preservation of  these primary constraints implies
 secondary constraints that together with primary constraints are second class constraints. Hence they can be explicitly solved with very interesting result. In more details, in order to find Hamiltonian formulation of the action proposed in  \cite{Bergshoeff:2018yvt} we should find metric inverse to the boost invariant metric  that defines string sigma model in stringy Newton-Cartan gravity. It turns out that crucial object for construction of such a metric is matrix valued Newton potential which is natural generalization of Newton potential defined in Newton-Cartan geometry. Then we will be able to find corresponding Hamiltonian and diffeomorphism constraints and we show that they are the first class constraints. As a next step we proceed to the solution of the second class constraints. It turns out that when we solve these constraints and insert this result into the original Hamiltonian constraint we find that the resulting constraint agrees with the Hamiltonian constraint found in \cite{Kluson:2018uss} which is very nice consistency check of both approaches. Note that the Hamiltonian found in \cite{Kluson:2018uss} was derived  with the help of the limiting procedure that defines Newton-Cartan geometry from the relativistic one \cite{Bergshoeff:2015uaa}. Finally we determine Lagrangian from corresponding Hamiltonian and we find that it agrees exactly with the Lagrangian found in \cite{Andringa:2012uz} which is again very nice consistency check.

Let us outline our results and suggest possible extension of this work. We find canonical structure of non-linear sigma model proposed recently in \cite{Bergshoeff:2018yvt}. We determine all constraints and we identify Hamiltonian and spatial diffeomorphism constraints and calculate Poisson brackets between them. We also determine second class constraints and perform their explicit solutions which gives the Hamiltonian constraint that agrees with the constraint found in \cite{Kluson:2018uss}.

The next important step in our canonical formulation of non-relativistic string theory is to perform analysis of T-duality since, as was shown in \cite{Bergshoeff:2018yvt}, T-duality of non-relativistic string theory is more complex than in case of its relativistic version. This analysis is currently in progress.

The structure of this paper is following. In the next section (\ref{second}) we review basic facts about stringy Newton-Cartan geometry and non-linear sigma model defined on it, following
\cite{Bergshoeff:2018yvt}. Then in section (\ref{third}) we perform canonical analysis of this theory. Finally in section (\ref{fourth}) we explicitly solve second class constraint and determine corresponding Hamiltonian.

\section{Non-Linear Sigma Model on Stringy  Newton-Cartan Geometry}\label{second}
In this section we define  stringy Newton-Cartan geometry, following
\cite{Bergshoeff:2018yvt}. Let $\mM$ is $D+1$ dimensional manifold
and let $\mathcal{T}_p$ is tangent space at point $p$. We decompose
$\mathcal{T}_p$ into longitudinal directions  indexed by $A=0,1$ and
transverse directions with $A'=2,\dots,d-1$. Two dimensional
foliation of $\mM$ is defined by generalized clock function
$\tau_\mu^{ \ A}$ that is also known as longitudinal vielbein field
that satisfies a constraint
\begin{equation}
D_\mu \tau_\nu^{ \ A}-D_\nu\tau_\mu^{ \ A}=0 \ ,
\end{equation}
where $D_\mu$ is covariant derivative with respect to the
longitudinal Lorentz transformations acting on index $A$. Let us
also introduce transverse vielbein field $E_\mu^{ \ A'}$. We further
introduce projective inverse $\tau^\mu_{ \ A}$ and $E^\mu_{ \ A'}$
that are defined as
\begin{eqnarray}
& &E_\mu^{ \ A'}E^\mu_{\ B'}=\delta^{A'}_{B'} \ , \quad
\tau^\mu_{ \ A}\tau_\mu^{ \ B}=\delta_A^{ B} \ , \quad
\tau_\mu^{ \ A}\tau^\nu_{ \ A}+E_\mu^{ \ A'}E^\nu_{ \ A'}=\delta^\nu_\mu \ ,
\nonumber \\
& & \tau^\mu_{ \ A}E_\mu^{ \ A'}=0 \ , \quad \tau_\mu^{ \ A}E^\mu_{ \ A'}=0 \ .
\nonumber \\
\end{eqnarray}
Let $\Sigma_A^{ \ A'}$ is a parameter of string Galilei boost transformations. Then
various components of NC geometry transform in the following way
\begin{eqnarray}
& &\delta_\Sigma \tau_\mu^{ \ A}=0 \ , \quad \delta_\Sigma E_\mu^{ \ A'}=-\tau_\mu^{ \ A}\Sigma_A^{ \ A'} \ , \nonumber \\
& &\delta_\Sigma \tau^\mu_{ \ A}=E^\mu_{ \ A'}\Sigma_A^{ \ A'}
\ , \quad \delta_\Sigma E^\mu_{ \ A'}=0 \ .
\nonumber \\
\end{eqnarray}
From vielbein we can construct longitudinal metric
$\tau_{\mu\nu}=\tau_\mu^{ \ A}\tau_\nu^{ \ B}\eta_{AB}$ and
transverse metric $h^{\mu\nu}=E^\mu_{ \ A'}E^\nu_{ \
B'}\delta^{A'B'}$ that are invariant under string Galilean boost
transformations.

It is clear that in order to define string moving in stringy Newton-Cartan background we
need transverse tensor $H_{\mu\nu}$ that is invariant under the string Galilei boost.
It turns out that this can be done when we introduce gauge field $m_\mu^{ \ A}$
and we can construct boost invariant tensor
\begin{equation}
H_{\mu\nu}=E_\mu^{ \ A'}E_\nu^{ \ B'}\delta_{A'B'}+(\tau_\mu^{ \ A}m_\nu^{ \ B}+
\tau_\nu^{ \ A}m_\mu^{ \ B})\eta_{AB} \ .
\end{equation}
In conclusion, $\tau_\mu^{ \ A}, E_\mu^{ \ A'}$ and $m_\mu^{ \ A}$ defines stringy Newton-Cartan geometry.

Now we are ready to proceed to the string sigma model that was  introduced in  \cite{Bergshoeff:2018yvt}. An important point is that this model is relativistic on two-dimensional world-sheet and hence it should be defined on the Riemann surface $\Sigma$.
 It turns out that this action contains
world-sheet scalars $x^\mu$ that parameterize an embedding string into target space time together with two worlds-sheet scalars that we denote as $\lambda$ and $\blambda$. These fields are needed for the realization of string Galilei symmetry on the world-sheet theory.

Now we will be more explicit. Let $\sigma^\alpha, \alpha=0,1$ parameterize world-sheet surface $\Sigma$. The sigma model is endowed with two dimensional world-sheet metric $\gamma_{\alpha\beta}$ and we introduce two dimensional vielbein
$e_\alpha^{ \ a} \ , a=0,1$ so that
\begin{equation}
\gamma_{\alpha\beta}=e_\alpha^{ \ a}e_\beta^{ \ b}\eta_{ab} \ ,
\end{equation}
where $\eta_{ab}=\mathrm{diag}(-1,1)$.
Using light-cone coordinates for the flat index $a$ on the world-sheet tangent space we define
\begin{equation}
e_\alpha\equiv e_\alpha^{ \ 0}+e_\alpha^{\ 1} \ , \quad
\bar{e}_\alpha\equiv e_\alpha^{ \ 0}-e_\alpha^{\  1} \ .
\end{equation}
We can also use light-cone coordinates for the flat index $A$ on the space-time
tangent space $\mathcal{T}_p$ and define
\begin{equation}
\tau_\mu\equiv \tau_\mu^{ \ 0}+\tau_\mu^{ \ 1} \ , \quad
\bar{\tau}_\mu=\tau_\mu^{ \ 0}-\tau_\mu^{\ 1} \ .
\end{equation}
Then we are ready to write sigma model for non-relativistic string on an arbitrary
string Newton-Cartan geometry,
nonrelativistic Kalb-Ramond B-field $B_{\mu\nu}$ and dilaton field $\phi$
 in the form \cite{Bergshoeff:2018yvt}
\begin{eqnarray}\label{Saction}
& &S=-\frac{T}{2}\int d^2\sigma (\sqrt{-\gamma}\gamma^{\alpha\beta}
\partial_\alpha x^\mu\partial_\beta x^\nu H_{\mu\nu}+\epsilon^{\alpha\beta}
(\lambda e_\alpha \tau_\mu+\bar{\lambda}\bar{e}_\alpha \bar{\tau}_\mu)
\partial_\beta x^\mu)-\nonumber \\
& &-\frac{T}{2}\int d^2\sigma \epsilon^{\alpha\beta}\partial_\alpha x^\mu
\partial_\beta x^\nu B_{\mu\nu}+\frac{1}{4\pi}\int d^2\sigma \sqrt{-h}R\phi \ ,
\nonumber \\
\end{eqnarray}
where $\gamma=\det \gamma_{\alpha\beta} \ , \gamma^{\alpha\beta}$ is inverse to
$\gamma_{\beta\alpha}$, $R$ is scalar curvature of $\gamma_{\alpha\beta}$ and $T$ is string tension.  Further, $\partial_\alpha x^\mu=\frac{\partial}{\partial \sigma^\alpha}x^\mu$.
In what follows we restrict to the case of constant dilaton field so that the
last term on the second line in (\ref{Saction}) is total derivative and will be ignored. It is important to stress that $\lambda$ and $\blambda$ are world-sheet scalars under change of wold-sheet coordinates $\sigma'^\alpha(\sigma)$. Explicitly, under such transformations we have
\begin{eqnarray}
\gamma'_{\alpha\beta}(\sigma')=\frac{\partial \sigma^\gamma}{\partial \sigma'^\alpha}
\frac{\partial \sigma^\delta}{\partial \sigma'^\beta}
\gamma_{\gamma\delta}(\sigma) \ , \quad \lambda'(\sigma')=\lambda(\sigma) \ , \quad
\blambda'(\sigma')=\blambda(\sigma) \ , \quad x'^\mu(\sigma')=x^\mu(\sigma) \ .
\nonumber \\
\end{eqnarray}
Further,  $\epsilon^{\alpha\beta}$ is Levi-Chivita symbol  defined as $\epsilon^{01}=-\epsilon^{10}=1$.

After this review of string sigma model in stringy Newton-Cartan background
we now proceed to its Hamiltonian formulation.
\section{Hamiltonian Formulation of String in Stringy Newton-Cartan Background}\label{third}
The presence of two dimensional vielbeins $e_\alpha^{ \ a}$ makes
the analysis slightly complicated and hence it is important to
choose suitable parametrization. To do this we use convention
introduced in \cite{Fujiwara:1996vp,Fujiwara:1995ys}. Explicitly,
let us define $e_\alpha$ and $\bar{e}_\alpha$ as
\begin{equation}
e_\alpha^{ \ 0}=\frac{1}{2}(e_\alpha+\bar{e}_\alpha) \ ,
\quad e_\alpha^{\ 1}=\frac{1}{2}(e_\alpha-\bar{e}_\alpha) \ .
\end{equation}
Then it is easy to see that $\gamma_{\alpha\beta}=e_\alpha^{ \ a}e_\beta^{ \ b}\eta_{ab}$ has the form
\begin{equation}
\gamma_{\alpha\beta}=-\frac{1}{2}(e_\alpha \bar{e}_\beta+
\bar{e}_\alpha e_\beta)
\end{equation}
and also  $\gamma=\det \gamma_{\alpha\beta}$ is equal to
\begin{equation}
\gamma=-\frac{1}{4}(e_0 \bar{e}_1-\bar{e}_0e_1)^2 \ .
\end{equation}
Then  inverse metric $\gamma^{\alpha\beta}$ has components
\begin{equation}
\gamma^{00}=\frac{4e_1 \bar{e}_1}{(e_0 \bar{e}_1-\bar{e}_0 e_1)^2} \ ,
\quad \gamma^{11}=\frac{4 e_0\bar{e}_0}{(e_0 \bar{e}_1-\bar{e}_0 e_1)^2} \ , \quad
\gamma^{01}=-2\frac{e_0 \bar{e}_1+e_1 \bar{e}_0}{(e_0 \bar{e}_1-\bar{e}_0e_1)^2} \ .
\end{equation}
As the next step we introduce following variables \cite{Fujiwara:1996vp,Fujiwara:1995ys}
\begin{eqnarray}
\xi=\ln (-e_1\bar{e}_1) \ , \quad \epsilon=\frac{1}{2}\ln \left(-\frac{e_1}{\bar{e}_1}\right) \ ,
\quad
\Gamma^+=\frac{e_0}{e_1} \ , \quad \Gamma^-=-\frac{\bar{e}_0}{\bar{e}_1}
\nonumber \\
\end{eqnarray}
with following inverse relation
\begin{eqnarray}
e_1&=&e^{\frac{1}{2}(\xi+2\epsilon)} \ , \quad
\bar{e}_1=e^{\frac{1}{2}(\xi-2\epsilon)} \ , \nonumber \\
e_0&=&\Gamma^+ e^{\frac{1}{2}(\xi+2\epsilon)} \ , \quad
\bar{e}_0=-\Gamma^-e^{\frac{1}{2}(\xi-2\epsilon)} \  \nonumber \\
\end{eqnarray}
and hence we obtain
\begin{eqnarray}
\sqrt{-\gamma}\gamma^{00}=-\frac{2}{\Gamma^++\Gamma^-}  ,
\quad
\sqrt{-\gamma}\gamma^{11}=\frac{2\Gamma^+\Gamma^-}{\Gamma^++\Gamma^-} \ ,
\quad
\sqrt{-\gamma}\gamma^{01}=\frac{\Gamma^+-\Gamma^-}{\Gamma^++\Gamma^-} \ .
\nonumber \\
\end{eqnarray}
With the help of these relations we rewrite the action (\ref{Saction}) into the form
\begin{eqnarray}\label{Sactionp}
S&=&T\int d^2\sigma (\frac{1}{\Gamma^++\Gamma^-}
\dot{x}^\mu-\Gamma^+x'^\mu)(\dot{x}^\nu+\Gamma^-x'^\nu)H_{\mu\nu}
-\nonumber \\
&-&\frac{T}{2}\int d^2\sigma
(\lambda \Gamma^+e^{\frac{1}{2}(\xi+2\epsilon)}\tau_\mu x'^\mu
    -\lambda e^{\frac{1}{2}(\xi+2\epsilon)}\tau_\mu\dot{x}^\mu)
    -\nonumber \\
&-&\frac{T}{2}\int d^2\sigma
(-\bar{\lambda}\Gamma^-e^{\frac{1}{2}(\xi-2\epsilon)}\bar{\tau}_\mu x'^\mu-
\bar{\lambda}e^{\frac{1}{2}(\xi-2\epsilon)}\bar{\tau}_\mu\dot{x}^\mu)
-\nonumber \\
&-&T\int d^2\sigma \dot{x}^\mu x'^\nu B_{\mu\nu} \ ,
\nonumber \\
\end{eqnarray}
where $\dot{x}^\mu\equiv \frac{\partial x^\mu}{\partial \sigma^0} \ ,x'^\mu=\frac{\partial x^\mu}{\partial \sigma^1}$. From the form of the action (\ref{Sactionp}) we see that it is natural to perform rescaling of  $\lambda, \bar{\lambda}$  as
\begin{equation}
\lambda^+=\lambda e^{\frac{1}{2}(\xi+2\epsilon)} \ , \quad
\lambda^-=\lambda e^{\frac{1}{2}(\xi-2\epsilon)}
\end{equation}
and hence the action (\ref{Sactionp}) has the form
\begin{eqnarray}\label{Sactionf}
S&=&T\int d^2\sigma \frac{1}{\Gamma^++\Gamma^-}
(\dot{x}^\mu-\Gamma^+x'^\mu)(\dot{x}^\nu+\Gamma^-x'^\nu)H_{\mu\nu}
-T\int d^2\sigma \dot{x}^\mu x'^\nu B_{\mu\nu}
-\nonumber \\
&-&\frac{T}{2}\int d^2\sigma
(\lambda^+ \Gamma^+\tau_\mu x'^\mu
-\lambda^+\tau_\mu\dot{x}^\mu)
+\frac{T}{2}\int d^2\sigma
(\lambda^-\Gamma^-\bar{\tau}_\mu x'^\mu+\lambda^-
\bar{\tau}_\mu\dot{x}^\mu) \ .
 \nonumber \\
\end{eqnarray}
Before we proceed to the canonical formalism we would like to
analyze the action (\ref{Sactionf}) in more details. Let us
determine equation of motion for $\lambda^+$ and $\lambda^-$ that
follow from (\ref{Sactionf})
\begin{eqnarray}
\Gamma^+\tau_\mu x'^\mu-\tau_\mu \dot{x}^\mu=0 \ , \quad
\Gamma^-\bar{\tau}_\mu x'^\mu+\bar{\tau}_\mu \dot{x}^\mu=0 \ .
\nonumber \\
\end{eqnarray}
If we combine these equations together we obtain
\begin{equation} \label{Gamma+-}
\Gamma^+\Gamma^-=-\frac{\tau_{\tau\tau}}{\tau_{\sigma\sigma}} \ ,  \quad
\Gamma^+-\Gamma^-=2\frac{\tau_{\tau\sigma}}{\tau_{\sigma\sigma}} \ ,
\end{equation}
where
\begin{equation}
\tau_{\alpha\beta}=\tau_{\mu\nu}\partial_\alpha x^\mu \partial_\beta x^\nu \ ,  \quad
\tau_{\mu\nu}=\tau_\mu^{ \ A}\tau_\nu^{ \ B}\eta_{AB} \ .
\end{equation}
Equations (\ref{Gamma+-}) can be solved as
\begin{equation}
\Gamma^-=\frac{-\tau_{\tau\sigma}+\sqrt{-\det\tau}}{\tau_{\sigma\sigma}} \ , \quad
\Gamma^+=\frac{\tau_{\tau\sigma}+\sqrt{-\det\tau}}{\tau_{\sigma\sigma}}  \ .
\end{equation}
Inserting this result into the action (\ref{Sactionf}) we finally  obtain
\begin{equation}
S=-\frac{T}{2}\int d^2\sigma \sqrt{-\det\tau}
\tau^{\alpha\beta}H_{\mu\nu}\partial_\alpha x^\mu
\partial_\beta x^\nu-T\int d^2\sigma \dot{x}^\mu x'^\mu B_{\mu\nu}
\end{equation}
which corresponds to the non-relativistic string action as was formulated
in \cite{Andringa:2012uz}.

Let us now proceed to the canonical formalism. From
(\ref{Sactionf}) we  obtain following conjugate momenta
\begin{equation}\label{pmu}
p_\mu=T\frac{1}{\Gamma^++\Gamma^-}
(2\dot{x}^\nu+(\Gamma^--\Gamma^+)x'^\nu)H_{\nu\mu}-TB_{\mu\nu}x'^\nu+\frac{T}{2}
\lambda^+\tau_\mu+\frac{T}{2}\lambda^-\bar{\tau}_\mu \ , \nonumber \\
\end{equation}
or equivalently
\begin{equation}
\Pi_\mu=\frac{2T}{\Gamma^++\Gamma^-}H_{\mu\nu}\dot{x}^\nu \ ,  \quad
\Pi_\mu=p_\mu-T\frac{\Gamma^--\Gamma^+}{\Gamma^++\Gamma^-}H_{\mu\nu}x'^\nu+TB_{\mu\nu}
x'^\nu-\frac{T}{2}\lambda^+\tau_\mu-\frac{T}{2}\lambda^-\bar{\tau}_\mu \ .
\nonumber \\
\end{equation}
Remaining conjugate momenta are primary constraints of the theory
\begin{eqnarray}
& &p^\Gamma_+=\frac{\partial L}{\partial \dot{\Gamma}^+}\approx 0 \ ,  \quad
p^\Gamma_-=\frac{\partial L}{\partial \dot{\Gamma}^-}\approx 0 \ ,
\nonumber \\
& &p_+^\lambda=\frac{\partial L}{\partial \dot{\lambda}^+}\approx 0 \ ,
\quad
p_-^\lambda=\frac{\partial L}{\partial \dot{\lambda}^-}\approx 0 \ .
\nonumber \\
\end{eqnarray}
Now using (\ref{pmu}) we obtain Hamiltonian density in the form
\begin{eqnarray}\label{mHnoncan}
\mH&=&p_\mu\dot{x}^\mu-\mL \nonumber \\
&=&\frac{T}{\Gamma^++\Gamma^-}(\dot{x}^\mu H_{\mu\nu}\dot{x}^\nu+\Gamma^+\Gamma^-
H_{\mu\nu}x'^\mu x'^\nu)+
\frac{T}{2}\lambda^+\Gamma^+\tau_\mu x'^\mu-\frac{T}{2}\lambda^- \Gamma^-\tau_\mu x'^\mu \ .
\nonumber \\
\end{eqnarray}
Of course, this is not correct form of the Hamiltonian density since
it does not depend on canonical variables $p_\mu,x^\mu$. In order to express it in the right
form we have to find relation between $\dot{x}^\mu$ and $p_\mu$. In order to solve
this problem  let us  observe that we have following relation
\begin{eqnarray}\label{Hh}
H_{\mu\rho}h^{\rho\sigma}H_{\sigma\nu}=
H_{\mu\nu}+\tau_\mu^{ \ A}\Phi_{AB}\tau_\nu^{ \ B} \ ,
\nonumber \\
\end{eqnarray}
where we defined matrix valued Newton potential $\Phi_{AB}$ as
\begin{eqnarray}
\Phi_{AB}=-\tau^\sigma_{ \ A}m_\sigma^{ \ C}\eta_{CB}
-\eta_{AC}m_\rho^{ \ C}\tau^\rho_{ \ B}
+\eta_{AC}m_\rho^{ \ C}h^{\rho\sigma}m_\sigma^{ \ D}\eta_{DB} \ .
\nonumber \\
\end{eqnarray}
Let us further define $\htau^\mu_{ \ A}$ as
\begin{equation}
\htau^\mu_{ \ A}=\tau^\mu_{ \ A}-h^{\mu\rho}m_\rho^{ \ B}\eta_{BA} \ .
\end{equation}
Then it is easy to see that
\begin{eqnarray}\label{tauPi}
\htau^\mu_{ \ A}\Pi_\mu=-\frac{2T}{\Gamma^++\Gamma^-}\Phi_{AB}\tau_\nu^{ \ B}\dot{x}^\nu \ .
\nonumber \\
\end{eqnarray}
To proceed further we will presume that $\Phi_{AB}$ is non-singular matrix so that
we can introduce its inverse in the form
\begin{equation}
(\Phi^{-1})^{AB}=\frac{1}{\det\Phi_{AB}}\left(\begin{array}{cc}
\Phi_{11} & -\Phi_{01} \\
-\Phi_{01} & \Phi_{00} \\ \end{array}\right) \ .
\end{equation}
Now if we combine (\ref{Hh}) with (\ref{tauPi}) we find that the inverse metric $H^{\mu\nu}$
to $H_{\mu\nu}$ has the form
\begin{equation}
H^{\mu\nu}\equiv
h^{\mu\nu}-\htau^\mu_{ \ A}(\Phi^{-1})^{AB}\htau^\nu_{ \ B} \ ,
\quad
H^{\mu\nu}H_{\nu\rho}=\delta^\mu_\rho  \ .
\end{equation}
then it is easy to determine canonical Hamiltonian from (\ref{mHnoncan}) and we obtain
\begin{eqnarray}
& &H=\int d\sigma \mH \ , \quad \mH=
\frac{(\Gamma^++\Gamma^-)}{4T}\pi_\mu
H^{\mu\nu}\pi_\nu+
\frac{T}{4}(\Gamma^++\Gamma^-)x'^\mu H_{\mu\nu}x'^\nu
-
\nonumber \\
& &-\frac{1}{2}(\Gamma^--\Gamma^+)x'^\mu \pi_\mu
-\frac{1}{4}(\Gamma^++\Gamma^-)\pi_\mu H^{\mu\nu}
(\lambda^+\tau_\nu+\lambda^-\bar{\tau}_\nu)+
\nonumber \\
& &+\frac{T}{4}(\Gamma^-+\Gamma^+)x'^\mu(\lambda^+\tau_\mu-
\lambda^-\bar{\tau}_\mu)+\frac{T}{16}(\Gamma^++\Gamma^-)(\lambda^+\tau_\mu+\lambda^-\bar{\tau}_\mu)
H^{\mu\nu}(\lambda^-\tau_\nu+\bar{\lambda}^-\tau_\nu) \ ,
\nonumber \\
\end{eqnarray}
where
\begin{equation}
\pi_\mu=p_\mu+TB_{\mu\rho}x'^\rho \ .
\end{equation}
Finally we introduce two variables $N$ and $N^\sigma$ defined as
\begin{equation}
N=\frac{1}{4}(\Gamma^++\Gamma^-) \ , \quad
N^\sigma=\frac{1}{2}(\Gamma^+-\Gamma^-)
\end{equation}
so that we find final form of the Hamiltonian density
\begin{equation}
\mH=N\mH_\tau+N^\sigma \mH_\sigma \ ,
\end{equation}
where
\begin{eqnarray}\label{defmHtau}
& &\mH_\tau=
\frac{1}{T}
\pi_\mu H^{\mu\nu}\pi_\nu+
Tx'^\mu H_{\mu\nu}x'^\nu-\pi_\mu H^{\mu\nu}
(\lambda^+\tau_\nu+\lambda^-\bar{\tau}_\nu)+
\nonumber \\
& &+Tx'^\mu(\lambda^+\tau_\mu-
\lambda^-\bar{\tau}_\mu)
+\frac{T}{4}(\lambda^+\tau_\mu+\lambda^-\bar{\tau}_\mu)
H^{\mu\nu}(\lambda^-\tau_\nu+\bar{\lambda}^-\tau_\nu) \ ,
\nonumber \\
& & \mH_\sigma=
x'^\mu p_\mu\ . \nonumber \\
\end{eqnarray}
Let us now proceed to the analysis of the requirement of the preservation
of all primary constraints. In case of the constraints
$p_N\approx 0 \ , p_\sigma\approx 0$ which are momenta conjugate to $N$ and $N^\sigma$ we obtain
\begin{eqnarray}
\dot{p}_N=\pb{p_N,H}=-\mH_\tau \approx 0 \ , \nonumber \\
\dot{p}_\sigma=\pb{p_\sigma,H}=-\mH_\sigma \approx 0 \ ,
\nonumber \\
\end{eqnarray}
while requirement of the preservation of the constraints $p_+^\lambda \approx 0 \ , p_-^\lambda\approx 0$
implies
\begin{eqnarray}
\dot{p}_+^\lambda=\pb{p_+^\lambda,H}=\pi_\mu H^{\mu\nu}\tau_\nu-
Tx'^\mu \tau_\mu-\frac{T}{2}\tau_\mu H^{\mu\nu}(\lambda^-\tau_\nu+\lambda^-\tau_\nu)\equiv
\mG^\lambda_+\approx 0 \ , \nonumber \\
\dot{p}_-^\lambda=\pb{p_-^\lambda,H}=\pi_\mu H^{\mu\nu}\bar{\tau}_\nu+
Tx'^\mu\bar{\tau}_\mu-\frac{T}{2}\bar{\tau}_\mu H^{\mu\nu}
(\lambda^-\tau_\nu+\blambda^-\bar{\tau}_\nu)\equiv \mG_-^\lambda\approx 0 \ .
\nonumber \\
\end{eqnarray}
Let us now analyze constraints $\mH_\sigma\approx 0 \ ,
\mH_\tau\approx 0$ in more details. Since we can anticipate that
$\mH_\sigma\approx 0$ is generator of spatial diffeomorphism it is
natural to extend it in the following way
\begin{equation}
\mH_\sigma \rightarrow p_\mu  x'^\mu+\lambda'^+ p^\lambda_++
\lambda'^- p^\lambda_-
\end{equation}
and introduce its smeared form
\begin{equation}
\bT_\sigma(N^\sigma)=\int d\sigma N^\sigma \mH_\sigma
\end{equation}
together with smeared form of the Hamiltonian constraint $\bT_\tau(N)=\int d\sigma N\mH_\tau$.  Note that $\bT_\sigma(N^\sigma)$ has non-zero Poisson bracket with canonical
variables
\begin{eqnarray}
\pb{\bT_\sigma(N^\sigma),x^\mu}=-N^\sigma  x'^\mu  \ , \quad
\pb{\bT_\sigma(N^\sigma),p_\mu}=- (N^\sigma p_\mu)' \ ,
\nonumber \\
\pb{\bT_\sigma(N^\sigma),\lambda^\pm}=-N^\sigma  \lambda'^\pm \ ,  \quad
\pb{\bT_\sigma(N^\sigma),p^\lambda_\pm}=- (N^\sigma p^\lambda_\pm)'
\ . \nonumber \\
\end{eqnarray}
Then it is easy to see that
\begin{eqnarray}
\pb{\bT_\sigma(N^\sigma),\bT_\sigma(M^\sigma)}=\bT_\sigma(N^\sigma M'^\sigma-
M^\sigma N'^\sigma) \ . \nonumber \\
\end{eqnarray}
In the same way we obtain
\begin{eqnarray}
\pb{\bT_\sigma(N^\sigma),\mH_\tau}=-2 N'^\sigma \mH_\tau-N^\sigma
\mH'_\tau \nonumber \\
\end{eqnarray}
or equivalently
\begin{equation}
\pb{\bT_\sigma(N^\sigma),\bT_\tau(M)}=\bT_\tau(N^\sigma M'-M N'^\sigma) \ .
\end{equation}
Finally we calculate Poisson bracket
\begin{eqnarray}
& &\pb{\bT_\tau(N),\bT_\tau(M)}=\int d\sigma (N M'-MN')(
p_\mu x'^\mu-2x'^\mu(\lambda^+\tau_\mu+\lambda^-\bar{\tau}_\mu)+\nonumber \\
& &+2\pi_\mu H^{\mu\nu}(\lambda^+\tau_\nu-\bar{\lambda}^-\bar{\tau}_\nu))
-T(
\lambda^+\tau_\mu-\bar{\lambda}^-\bar{\tau}_\mu)H^{\mu\nu}(\lambda^+\tau_\nu+
\lambda^-\bar{\tau}_\nu)=
\nonumber \\
& &=\int d\sigma (NM'-M N')
(\mH_\sigma+2(\lambda^+\mG_+^\lambda-\lambda^-\mG_-^\lambda))
\nonumber \\
\end{eqnarray}
that vanishes on the constraint surface $\mH_\sigma\approx 0 \ , \mG_+^\lambda\approx 0 \ ,
\mG_-^\lambda \approx 0$. Collecting all these results together we find
that $\mH_\tau\approx 0 \ ,  \mH_\sigma\approx 0$ are the first class constraints
which is an expected result since the action (\ref{Saction}) defines relativistic
theory on two-dimensional world-sheet $\Sigma$.
\section{Second Class Constraints and Their Solution}\label{fourth}
Now we analyze the constraints $\mG_+^\lambda\approx 0 \ ,
\mG_-^\lambda\approx 0$ in more details. First of all we show that
 $\mG_+^\lambda\approx 0 \ , \mG_-^\lambda \approx 0 $ are second class constraints together with $p^\lambda_+\approx 0$ and $p^\lambda_-\approx 0$  since
\begin{eqnarray}
& &\pb{p^\lambda_+(\sigma),\mG^\lambda_+(\sigma')}=\frac{T}{2}\tau_\mu H^{\mu\nu}\tau_\nu
\delta(\sigma-\sigma') \ , \quad
\pb{p^\lambda_+(\sigma),\mG^\lambda_-(\sigma')}=\frac{T}{2}\bar{\tau}_\mu H^{\mu\nu}\tau_\nu \delta(\sigma-\sigma')\ , \nonumber \\
& &\pb{p^\lambda_-(\sigma),\mG^\lambda_+(\sigma')}=\frac{T}{2}\tau_\mu H^{\mu\nu}\bar{\tau}_\nu
\delta(\sigma-\sigma') \ , \quad
\pb{p^\lambda_-(\sigma),\mG^\lambda_-(\sigma')}=\frac{T}{2}
\bar{\tau}_\mu H^{\mu\nu}\bar{\tau}_\nu\delta(\sigma-\sigma') \ .
\nonumber \\
\end{eqnarray}
Clearly there is also non-zero Poisson bracket between $\mG_+^\lambda\approx 0$ and
$\mG_-^\lambda\approx 0$. Let us now introduce common notation for the second class constraint as $\Psi_A\equiv (p^\lambda_+,p^\lambda_-,\mG^\lambda_+,\mG^\lambda_-)$. Then the matrix of Poisson brackets between these constraints has schematic form
\begin{equation}
\triangle_{AB}=\left(\begin{array}{cc}
0 & X \\
Y & W \\ \end{array}\right) \ ,
\end{equation}
where $X,Y,W$ are $2\times 2$ matrices that have generally inverse
matrices \footnote{Of course, each entry of these matrices is
infinite dimensional since it depends generally on $\sigma$ and
$\sigma'$. However for our purposes this schematic form is
sufficient.}. Then the inverse matrix has the form
\begin{equation}
\triangle^{AB}=\left(\begin{array}{cc}
-Y^{-1}W X^{-1} & Y^{-1} \\
X^{-1} & 0 \end{array}\right) \ , \quad  \triangle_{AB}\triangle^{BC}=\delta_A^C \ .
\end{equation}
If we now calculate Dirac bracket between $x^\mu$ and $p_\nu$ we obtain
\begin{eqnarray}
& &\pb{x^\mu,p_\nu}_D=\pb{x^\mu,p_\nu}-\pb{x^\mu,\Psi_A}\triangle^{AB}
\pb{\Psi_B,p_\nu}=\nonumber \\
& & =\pb{x^\mu,p_\nu}-(0,0,*,*)
\left(\begin{array}{cc}
-Y^{-1}W X^{-1} & Y^{-1} \\
X^{-1} & 0 \end{array}\right)(0,0,*,*)^T=\pb{x^\mu,p_\nu} \ ,
\nonumber \\
\end{eqnarray}
where $*$ means non-zero entry whose explicit form is not important.
From this result we see that Dirac brackets between
$x^\mu$ and $p_\nu$ coincide with corresponding Poisson brackets. Now we are ready to solve
 the second class constraints $\mG^\lambda_+\approx 0$ and
$\mG^\lambda_-\approx 0$. First of all we introduce part of the Hamiltonian constraint
$\mH_\tau\approx 0$ that depends on $\lambda^+$ and $\lambda^-$ as
\begin{equation}
\mH^\lambda_\tau=A\lambda^++B\lambda^-+\frac{T}{4}((\lambda^+)^2X+
2\lambda^+\lambda^-Y+(\lambda^-)^2W) \ ,
\end{equation}
where
\begin{eqnarray}\label{defAB}
& &A=-\pi_\mu H^{\mu\nu}\tau_\nu+Tx'^\mu\tau_\mu \ , \quad
B=-\pi_\mu H^{\mu\nu}\bar{\tau}_\nu-T x'^\nu\bar{\tau}_\nu \ , \nonumber \\
& & X=\tau_\mu H^{\mu\nu}\tau_\nu \ , \quad Y=\bar{\tau}_\mu H^{\mu\nu}\tau_\nu \ ,
\quad W=\bar{\tau}_\mu H^{\mu\nu}\bar{\tau}_\nu \ . \nonumber \\
\end{eqnarray}
Using this notation we can write the solution of the second class constraints
$\mG_+^\lambda=0, \quad  \mG_-^\lambda=0$ in the form
\begin{eqnarray}
\lambda^-=-\frac{2}{T}\frac{AY-BX}{Y^2-XW} \ , \quad
\lambda^+=-\frac{2}{T}\frac{BY-AW}{Y^2-XW} \ . \nonumber \\
\end{eqnarray}
Then inserting this result into $\mH^\lambda_\tau$ we obtain
\begin{eqnarray}
\mH^\lambda_\tau(on shell)
=\frac{1}{T(Y^2-XW)}
(A^2W+B^2X-2ABY) \ ,
\nonumber \\
\end{eqnarray}
where explicit calculations give
\begin{eqnarray}
X&=&
-(\Phi^{-1})^{00}-2(\Phi^{-1})^{01}-(\Phi^{-1})^{10} \ ,
\quad
Y=
(\Phi^{-1})^{AB}\eta_{BA} \ , \nonumber \\
W&=&
-(\Phi^{-1})^{00}+2(\Phi^{-1})^{01}
-(\Phi^{-1})^{11} \  \nonumber \\
\end{eqnarray}
so that
\begin{eqnarray}
Y^2-XW
=-\frac{4}{\det \Phi_{AB}} \ .  \nonumber \\
\end{eqnarray}
Then after some complicated calculations and with the help
of the  explicit form of $A$ and $B$ given  in (\ref{defAB})  we get
\begin{eqnarray}\label{mHtaulambdaon}
& &\mH_\tau^\lambda(on shell)=
\nonumber \\
&=&\frac{1}{T}\pi_\mu \htau^\mu_{ \ A}(\Phi^{-1})^{AB}\htau^\nu_{ \ B}\pi_\nu
-2\pi_\mu\htau^\mu_{ \ A}\epsilon^{AB}\eta_{BC}\tau_\sigma^{ \ C}+T
\tau_\sigma^{ \ A}\Phi_{AB}\tau_\sigma^{ \ B}-T\tau_\sigma^{ \ A}\tau_\sigma^{ \ B}\eta_{AB}
\Phi_{CD}\eta^{CD} \ , \nonumber \\
\end{eqnarray}
where  $\tau_\sigma^{ \ A}\equiv x'^\mu \tau_\mu^{ \ A}$.
Inserting (\ref{mHtaulambdaon}) into (\ref{defmHtau})  we obtain
Hamiltonian constraint that depends on the canonical variables $x^\mu$ and $p_\mu$ only
\begin{eqnarray}\label{mHtausol}
& &\mH^{sol}_\tau=\frac{1}{T}\pi_\mu h^{\mu\nu}p_\nu
+Tx'^\mu H_{\mu\nu}x'^\nu
\nonumber \\
& & -2\pi_\mu \htau^\mu_{ \ A}\epsilon^{AB}\eta_{BC}\tau_\sigma^{ \ C}
+T
\tau_\sigma^{ \ A}\Phi_{AB}\tau_\sigma^{ \ B}-T\tau_\sigma^{ \ A}\tau_\sigma^{ \ B}\eta_{AB}
\Phi_{CD}\eta^{CD}
\ ,
\nonumber \\
\end{eqnarray}
where $\tau_\sigma^{ \ A}\equiv x'^\mu \tau_\mu^{ \ A}$.
The form of the Hamiltonian constraint (\ref{mHtausol})
  coincides with the Hamiltonian constraint
found in \cite{Kluson:2018uss} where non-relativistic string in stringy Newton-Cartan
background was defined with the help of the limiting procedure that defines
Newton-Cartan geometry from the relativistic one. We mean that this is very nice
consistency check of both approaches.

Finally we would like to check the analysis further and try to
determine corresponding Lagrangian density.  Using canonical
equation of motion we get
\begin{equation}
\dot{x}^\mu=\pb{x^\mu,H}=
\frac{2N}{T}h^{\mu\nu}\pi_\nu-2N\htau^\mu_{ \ A}\epsilon^{AB}
\eta_{BC}\tau_\sigma^{ \ C}+N^\sigma \partial_\sigma x^\mu \ ,
\end{equation}
where $H=\int d^\sigma (N\mH_\tau^{sol}+N^\sigma \mH_\sigma)$.
Then we find
\begin{eqnarray}\label{mLpi}
\mL&=&p_\mu\dot{x}^\mu-N\mH^{sol}_\tau-N^\sigma
\mH_\sigma=\nonumber \\
&=&\frac{N}{T}\pi_\mu h^{\mu\nu}\pi_\nu
-T
\tau_\sigma^{ \ A}\Phi_{AB}\tau_\sigma^{ \ B}+T\tau_\sigma^{ \ A}\tau_\sigma^{ \ B}
\Phi_{AB}
-Tx'^\mu H_{\mu\nu}x'^\nu
-T\partial_\tau x^\mu B_{\mu\nu}\partial_\sigma x^\nu \ .  \nonumber \\
\end{eqnarray}
To proceed further we will now follow
\cite{Kluson:2018uss} and  introduce $\hE_\mu^{ \ A'}$ defined as
\begin{equation}
\hE_\mu^{ \ A'}=E_\mu^{ \ A'}+m_\nu^{ \ A}E^\nu_{ \ C'}\delta^{C'A'}\tau_\mu^{ \ B}\eta_{BA}
\end{equation}
that obeys an important relation
\begin{equation}
\hE_\mu^{ \ A'}\htau^\mu_{ \ B}=0 \ .
\end{equation}
Then it is easy to express Lagrangian density (\ref{mLpi}) as function
of $x^\mu$ and $\partial_\alpha x^\mu$ and we obtain
\begin{eqnarray}\label{mLNNsigma}
\mL&=&\frac{T}{4N}
(\dot{x}^\mu-N^\sigma x'^\mu)\hE_\mu^{ \  A'}\delta_{A'B'}
\hE_\nu^{ \  B'}(\dot{x}^\nu-N^\sigma  x'^\nu)-\nonumber \\
&-&TN
\tau_\sigma^{ \ A}\Phi_{AB}\tau_\sigma^{ \ B}+TN\tau_\sigma^{ \ A}\tau_\sigma^{ \ B}\eta_{AB}
\Phi_{CD}\eta^{CD}
-TNx'^\mu H_{\mu\nu}x'^\nu
-T\partial_\tau x^\mu B_{\mu\nu}\partial_\sigma x^\nu \ . \nonumber \\
\nonumber \\
\end{eqnarray}
As the next step we determine Lagrange multipliers $N$ and $N^\sigma$. It turns out that
these multipliers are determined by equations of motion for $x^\mu$. In fact, if we multiply
this equation by $\tau_{\mu\nu}$ we obtain
\begin{equation}\label{taueqxmu}
\tau_{\mu\nu}(\dot{x}^\nu-N^\sigma x'^\nu)=
-2N\tau_\mu^{ \ E}\epsilon_{ED}\tau_\sigma^{ \ D} \ .
\end{equation}
If we further multiply this result with $ x'^\mu$ and use an antisymmetry of $\epsilon_{AB}$ we obtain
\begin{equation}\label{Nsigma}
N^\sigma=\frac{\tau_{\tau\sigma}}{\tau_{\sigma\sigma}} \ .
\end{equation}
If we manipulate with (\ref{taueqxmu}) further we  get
\begin{equation}
(\dot{x}^\mu-N^\sigma
 x'^\mu)\tau_{\mu\nu}(\dot{x}^\nu-N^\sigma
 x'^\nu)=-4N^2\tau_{\sigma\sigma}
\end{equation}
that can be written as
\begin{equation}
\tau_{\tau\tau}-2N^\sigma \tau_{\sigma\tau}+(N^\sigma)^2\tau_{\sigma\sigma}=
-4N^2\tau_{\sigma\sigma} \ .
\end{equation}
This equation can be solved for $N$ when we take into account the result
(\ref{Nsigma}) and we obtain
\begin{equation}\label{N}
N=\frac{1}{2}\frac{\sqrt{-\det\tau_{\alpha\beta}}}{\tau_{\sigma\sigma}} \ .
\end{equation}
Inserting (\ref{Nsigma}) and (\ref{N}) into (\ref{mLNNsigma})
we obtain Lagrangian density in the form
\begin{eqnarray}\label{mLamostdone}
\mL&=&-\frac{T}{2}\sqrt{-\det\tau_{\alpha\beta}}\tau^{\alpha\beta}
H_{\alpha\beta}-\frac{T}{2}\sqrt{-\det\tau_{\alpha\beta}}\tau^{\alpha\beta}
\tau_\alpha^{ \ A}\Phi_{AB}\tau_\beta^{ \ B}\nonumber \\
&+&\frac{T}{2}\sqrt{-\det\tau_{\alpha\beta}}
\Phi_{AB}\eta^{AB}-T\dot{x}^\mu B_{\mu\nu}x'^\nu  \ , \nonumber \\
\end{eqnarray}
where $H_{\alpha\beta}=H_{\mu\nu}\partial_\alpha x^\mu \partial_\beta x^\nu \ , \tau_\alpha^{ \ A}=\partial_\alpha x^\mu \tau_\mu^{ \ A}$ and where we used the fact that
\begin{equation}
\hE_\mu^{ \ A'}\delta_{A'B'}\hE_\nu^{ \ B'}=
H_{\mu\nu}+\tau_\mu^{ \ A}\Phi_{AB}\tau_\nu^{ \ B} \ .
\end{equation}
We see that this Lagrangian density almost coincides with the Lagrangian density
found \cite{Andringa:2012uz} up to terms that contain matrix valued Newton potential
$\Phi_{AB}$. Now we are going to argue that these terms cancel each other. In fact,
note that $\tau_{\alpha\beta}$ is defined as
\begin{equation}
\tau_{\alpha\beta}=\tau_\alpha^{ \ A}\tau_\beta^{ \ B}\eta_{AB} \ ,
\end{equation}
where $\tau_{\alpha}^{ \ A}$ is $2\times 2$ matrix. Now since $\tau_{\alpha\beta}$ is
non-singular so that $\tau_{\alpha}^{ \ A}$  is non-singular as well
and hence we can introduce
 an inverse matrix $\tau^\beta_{\ A}$ that obeys
the relation
\begin{equation}
\tau^\alpha_{ \ A}\tau_\alpha^{ \ B}=\delta_A^{ \ B} \ .
\end{equation}
Then we can
 define $\tau^{\alpha\beta}$ as
\begin{equation}
\tau^{\alpha\beta}=\tau^\alpha_{ \ A}\tau^\beta_{ \ B}\eta_{AB}
\end{equation}
that obeys
\begin{equation}
\tau^{\alpha\beta}\tau_\beta^{ \ A}=\tau^\alpha_{ \ C}\eta^{CA} \ ,
\end{equation}
and hence
\begin{equation}
\tau^{\alpha\beta}\tau_\beta^{ \ B}\tau_\alpha^{ \ A}=
\tau_\beta^{ \ B}\tau^\beta_{ \ C}\eta^{CA}=\eta^{BA} \ .
\end{equation}
With the help of these results it is  easy to see that contributions to the Lagrangian density (\ref{mLamostdone}) that depend on   $\Phi_{AB}$ cancel each
other
and hence the Lagrangian density has the final form
\begin{equation}
\mL=-\frac{T}{2}\sqrt{-\det\tau_{\alpha\beta}}\tau^{\alpha\beta}
H_{\alpha\beta}-T\dot{x}^\mu B_{\mu\nu} x'^\nu
\end{equation}
which is Lagrangian density proposed in \cite{Andringa:2012uz}. This result
again confirms validity of our approach.%

\acknowledgments{This  work  was
    supported by the Grant Agency of the Czech Republic under the grant
    P201/12/G028. }


\end{document}